\documentclass[a4paper,12pt]{article}

\usepackage[english]{babel}
\usepackage{graphics,amssymb,epsfig,float}
\usepackage{graphicx}% Include figure files
\usepackage{epsfig}% Include figure files
\usepackage{rotating}% Include figure files
\usepackage{dcolumn}% Align table columns on decimal point
\usepackage{bm}% bold math
\usepackage{cite}
\usepackage{amsmath}
\usepackage{enumerate}
\usepackage[usenames,dvipsnames]{xcolor}
\usepackage{color}
\usepackage[urlcolor=blue]{hyperref}
\usepackage[all]{hypcap}

\def\lsim{\;\raise0.5ex\hbox{\raise-0.1ex\hbox{$<$}\kern-0.75em\raise-1.1ex\hbox{$\sim$}}\;}
\def\gsim{\;\raise0.5ex\hbox{\raise-0.1ex\hbox{$>$}\kern-0.75em\raise-1.1ex\hbox{$\sim$}}\;}
\def\nn{\nonumber}

\def\a{\alpha}

\def\g{\gamma}
\def\d{\delta}
\def\t{\theta}
\def\k{\kappa}
\def\l{\lambda}
\def\s{\sigma}

\newcommand{\di}{\mathrm{d}}
\newcommand{\e}{\mathrm{e}}
\newcommand{\TeV}{\,\textrm{TeV}}
\newcommand{\GeV}{\,\textrm{GeV}}
\newcommand{\MeV}{\,\textrm{MeV}}

\newcommand{\eV}{\,\textrm{eV}}
\newcommand{\wt}{\widetilde}
\renewcommand{\=}{\!\!\!&=&\!\!\!}

\newcommand{\figref}[1]{\hyperref[#1]{Fig.\hspace{0.3ex}\ref*{#1}}}
\newcommand{\secref}[1]{\hyperref[#1]{Sec.\hspace{0.3ex}\ref*{#1}}}
\newcommand{\appref}[1]{\hyperref[#1]{App.\hspace{0.3ex}\ref*{#1}}}

\begin{document}

\begin{titlepage}

\begin{flushright}
LPT Orsay 13-55
\end{flushright}
\vspace{1cm}

\begin{center}
{\Large\bf Right-handed sneutrinos as asymmetric DM and}

\vspace{5mm}
{\Large\bf neutrino masses from neutrinophilic Higgs bosons} 

\vspace{2cm}

{\bf{Pantelis Mitropoulos}}
\vspace{6mm}\\
\textit{Laboratoire de Physique Th\'eorique, UMR 8627, CNRS and Universit\'e de Paris--Sud,}\\
\textit{B\^at. 210, 91405 Orsay, France}
\vspace{1mm}\\
Email: pantelis.mitropoulos@th.u-psud.fr

\end{center}

\vspace{1cm}

\begin{abstract}

We consider an extension of the Next-to-Minimal Supersymmetric Standard
Model by three  right-handed neutrinos and a pair of neutrinophilic
Higgs superfields. The small neutrino masses arise
naturally from a small vacuum expectation value of the additional
Higgs fields (hence without lepton number violation), while the lightest
right-handed  sneutrinos can constitute asymmetric Dark Matter.
The right-handed sneutrino and  baryon asymmetries are connected through
equilibrium processes in the early  universe, explaining the coincidence
of the DM and baryon abundances. We show that particle physics and
astrophysical constraints are satisfied.

\end{abstract}

\end{titlepage}

\section{Introduction}

A large part of the total matter of the universe manifests
itself only  through  its gravitational effects. Although the evidence
for its existence is strong, the nature  of this Dark Matter (DM) is not
yet clear. Observations of the Bullet Cluster  \cite{Clowe:2006eq}
show a clear separation between the location of the luminous  matter
and the location of gravitating matter providing a
confirmation of particle DM. Assuming the standard cosmological model
and using observations of the  cosmic microwave background, it is
possible to determine with great accuracy the present  abundance of
DM \cite{Hinshaw:2012aka,Ade:2013zuv}.

In most scenarios, the DM abundance is a relic of annihilation
processes  during the early hot universe. These processes stopped when
the expansion rate of  the universe became larger than their rate and,
since then, the DM density per comoving  volume has remained constant.
 In order that the relic density  fits the
observed value, the annihilation cross section should be roughly of the
order of weak interactions. Due to  this fact Weakly Interacting
Massive Particles (WIMPs) are preferred candidates  for~DM.

The Standard Model (SM) of particle physics does not include a particle
with the  required characteristics. Supersymmetric (R-parity
conserving) extensions of the Standard Model provide two stable
candidates, the neutralino and the sneutrino. However, even though
they are both weakly  interacting particles, their relic density does
not automatically have the correct value. In the case of neutralinos it
can vary  over many  orders of magnitude as a function of the unknown
parameters of the theory. (Left-handed) sneutrinos are problematic due
to their  strong coupling to the Z bosons. This strong coupling results
in a large annihilation cross section and hence in a too small  relic
density compared to the observed value, and in a scattering cross
section off  nuclei far above the current bounds from XENON100
\cite{Aprile:2012nq}.

The present baryonic matter abundance, on the other hand, is not
sensitive to the  baryon  pair annihilation cross section. It originates
from the baryon asymmetry, since all the anti-baryons have already been
completely depleted. Astonishingly, both DM and baryon  abundances are
of the same order of magnitude ($\Omega_{DM}/\Omega_b \simeq 5$).  This
coincidence of the values of these two quantities motivated the search for
alternative mechanisms for the generation of the DM relic
density, connected  to the one of the baryons (see 
\cite{Nussinov:1985xr,Barr:1990ca,Barr:1991qn,Dodelson:1991iv,
Kaplan:1991ah} for  early discussions and
\cite{Davoudiasl:2012uw,Petraki:2013wwa} for reviews). If the DM
particles carry a quantum number related to baryon number, the same
mechanism might be responsible for the generation of their relic density
via their asymmetry. 

At first sight, sneutrinos are promising candidates for asymmetric DM
(ADM)\cite{Hooper:2004dc,McDonald:2006if,Abel:2006nv,Page:2007sh,Kang:2011wb,
Ma:2011zm}\footnote{Higgsinos could also be ADM, but only if a number of strong
constraints is satisfied, see \cite{Blum:2012nf}.}. They carry a
conserved quantum number, lepton number, such that they can share
the asymmetry of charged leptons through processes which were in
equilibrium in the early hot universe. Then, their asymmetry can have
become related to the baryon asymmetry through sphaleron processes 
\cite{Harvey:1990qw,Chung:2008gv}. However, although the large
annihilation cross section between sneutrinos and anti-sneutrinos  is a
good feature allowing for ADM, their self-annihilation
(sneutrino--sneutrino or  anti-sneutrino--anti-sneutrino annihilation)
cross sections are also typically large,  destroying the 
asymmetry~\cite{Ellwanger:2012yg}.

However, non-zero neutrino masses suggest the existence of
right-handed neutrinos (and sneutrinos).
In the past years, a variety of models aiming at the
explanation of neutrino masses have been proposed (see, for
example,  \cite{Valle:2006vb,Morisi:2012fg} and references therein).
They can be categorized in  two main classes, those which employ
Majorana mass terms for right-handed neutrinos and  those employing
Dirac mass terms only. The former allow for various versions of  seesaw
mechanisms, amongst others the inverse seesaw which allows for
electroweak  scale right-handed neutrinos. The common characteristic of
these models is the violation of lepton number by a Majorana mass
term. Dirac neutrino
masses are less well studied, though not less motivated. The simplest 
way to obtain Dirac masses for neutrinos is the introduction of Yukawa
couplings to  Higgs bosons, but with unnatural small Yukawa coupling
constants. A more elegant way is  the introduction of an additional
Higgs field which couples only to right-handed  neutrinos. Then, the
smallness of neutrino masses is no longer due to small values of 
coupling constants, but can be due to a small vacuum expectation value
(vev) of the new  Higgs field.

The presence of right-handed sneutrinos opens new  possibilities
for sneutrino DM:
right-handed sneutrinos with a small left-handed component
may have at the same time a large pair annihilation cross section,
but a negligible self-annihilation cross section.
However, in scenarios as seesaw models which do not conserve lepton
number, an asymmetry of sneutrinos is difficult
to maintain due to oscillations between sneutrinos and anti-sneutrinos
\cite{Cirelli:2011ac,Buckley:2011ye,Tulin:2012re} (see also 
\cite{Arina:2011cu}).
In the Dirac case with small Yukawa couplings, asymmetric DM faces
the following difficulties:
First, the annihilation cross section, proportional to the
small couplings, is not adequate to eliminate the symmetric part of
the DM, resulting in a large relic density
unrelated to the asymmetry. Second, such small
Yukawa couplings keep the right-handed neutrinos and sneutrinos out
of equilibrium in the early universe and, as a
result, the asymmetry of the sneutrinos was never related to the
baryonic asymmetry.
However, if the small neutrino masses originate from the
small vev of an additional Higgs field (but with large Yukawa
couplings), these difficulties are solved.

Such scenarios, often denoted as \textit{neutrinophilic Higgs doublet
models}, appeared  first in \cite{Gabriel:2006ns,Wang:2006jy} (for an
earlier approach, but with Majorana  neutrinos, see \cite{Ma:2000cc}).
In these models, a $Z_2$ symmetry is  spontaneously broken generating a
small vev of the new Higgs scalar. This mechanism  results also in a
very light scalar with mass of the order of eV. Such light scalars  have
been ruled out \cite{Zhou:2011rc,Sher:2011mx} by astrophysical
arguments. However,  the $Z_2$ symmetry can be replaced by a global
$U(1)$ symmetry in order to forbid  Majorana masses, but which is
broken explicitly so that a very light scalar  is avoided
\cite{Davidson:2009ha}. The $U(1)$ symmetry makes
very small explicit breaking
terms (see below) more
natural. The  LHC phenomenology of this model is studied in
\cite{Davidson:2010sf}, while in  \cite{Marshall:2009bk} a
supersymmetric variant based on the MSSM and its  phenomenology are
examined. The additional Higgs doublets still allow the  SUSY version to
be embedded into a grand unified symmetry as verified in 
\cite{Haba:2012ai}. Furthermore, the additional Higgs doublets do not
spoil proton stability since  they couple only to leptons. In
\cite{Haba:2011fn}, the Higgs potential is studied for both SUSY and
non-SUSY models.  Scenarios for leptogenesis with neutrinophilic Higgs
are discussed in  \cite{Chao:2012pt,Haba:2011yc,Haba:2011ra}. Finally,
the sneutrino of the SUSY  model of \cite{Marshall:2009bk} has been used
as DM candidate in  \cite{Choi:2012ap,Choi:2013fva}. In particular, the
possibility of ADM is also considered in \cite{Choi:2013fva}, but with a
trilinear soft coupling of the order of  several TeV and a
(related) very large
annihilation cross section into monochromatic photons.

In this paper we consider the Next-to-Minimal Supersymmetric Standard
Model (NMSSM)  extended by a pair of neutrinophilic Higgs doublets and
three generations of  right-handed neutrino superfields.  The NMSSM
provides a natural solution of the  $\mu$-problem of the MSSM by
introducing a gauge singlet superfield $S$  (for reviews, see
\cite{Maniatis:2009re,Ellwanger:2009dp}). In the NMSSM, the Higgs mass
term $\mu$ in the superpotential of the MSSM is replaced by the term $\l
S H_u \cdot H_d$. Otherwise the singlet plays no particular role in our
model (its vev could be replaced by a constant dimensionful parameter),
but due to the additional coupling $\l$ the lightest CP-even Higgs is
naturally heavier  than in the MSSM
\cite{Ellis:1988er,Drees:1988fc,Ellwanger:1993hn,Ellwanger:2006rm}. 
Therefore, a SM-like Higgs mass of $\sim 125\GeV$ is much easier to
explain 
\cite{Hall:2011aa,Ellwanger:2011aa}.

We are going to explore whether and under which circumstances this model
can accommodate  right-handed sneutrinos as ADM. 
We find that this is
indeed possible under certain conditions. First, we note that the
ordinary Higgs sector of the NMSSM is not affected by the introduction
of the neutrophilic Higgses (henceforth $\nu$-Higgses). The scalar
$\nu$-Higgses, however, have to be relatively heavy of
${\cal{O}}(1)\TeV$ such that the additional degrees of freedom of the
Dirac  neutrinos do not lead to $^4\textrm{He}$ overabundance through
their contribution  to the expansion rate of the universe during big
bang nucleosynthesis (BBN). On the other  hand, light $\nu$-higgsinos are
required for a large sneutrino--anti-sneutrino pair annihilation cross
section which is necessary for  the sneutrino relic density to be
determined by its asymmetry.

In the next section we present the model and discuss constraints from
lepton number violation and BBN. In \secref{DM}, we explore  the
possibility for right-handed sneutrinos as asymmetric DM. In particular,
in \secref{LeptonNumber}
we examine  the connection between the sneutrino and baryon asymmetry
via sphaleron processes,
in \secref{SneutrinoADM}  we study conditions from oscillations, self
and pair annihilations, and in
\secref{detection} we discuss possible DM signals and constraints
from DM detection. 
Finally, we summarize our results in \secref{conclusions}.

\section{The model}\label{model}

We extend the NMSSM by three right-handed neutrino superfields
$\widehat{\nu}_R^c$ and  a pair of new Higgs doublets
$\widehat{H}_u^\nu$ and $\widehat{H}_d^\nu$. These fields  are charged
under a new global $U(1)$ symmetry with charges $-1$, $+1$ and $-1$,
respectively,  while the usual NMSSM superfields remain uncharged. The
superpotential is written as
\begin{equation} \label{eq:1}
 W=W^{NMSSM}+y_\nu \widehat{L}\cdot \widehat{H}_u^\nu \widehat{\nu}_R^c
+ \l_\nu  \widehat{S} \widehat{H}_u^\nu \cdot \widehat{H}_d^\nu,
\end{equation}
where the Yukawa coupling $y_\nu$ and the superfields $\widehat{L}$ and 
$\widehat{\nu}_R^c$ should be understood as matrix and vectors,
respectively,  in flavor space. The corresponding soft SUSY breaking
masses and couplings are
\begin{eqnarray} \label{eq:2}
 - \mathcal{L}_{soft} \= -\mathcal{L}_{soft}^{NMSSM} + m^2_{H_u^\nu}|
H_u^\nu |^2 +  m^2_{H_d^\nu}| H_d^\nu |^2 + m^2_{\nu_R} |\nu_R|^2
\nonumber \\
 && + y_\nu A_\nu L\cdot H_u^\nu \nu_R^c + \l_\nu A_{\l_\nu} S H_u^\nu
\cdot H_d^\nu.
\end{eqnarray}
$W^{NMSSM}$ and $\mathcal{L}_{soft}^{NMSSM}$ are the superpotential and
the soft terms  of the $\mathbb{Z}_3$-invariant NMSSM (see, e.g.,
\cite{Ellwanger:2009dp}),  respectively.

The new $U(1)$ symmetry needs to be broken by the vev of the $\nu$-Higgs
in order to give masses to the neutrinos. To this end we  add to the
Lagrangian 
\eqref{eq:2} the additional soft terms
\begin{displaymath}
 A_{\l_1} S H_u \cdot H_d^\nu + A_{\l_2} S H_u^\nu \cdot H_d.
\end{displaymath} These two terms do not correspond to terms in the
superpotential. Since they break  the $U(1)$ explicitly, it is natural
in the 't Hooft sense for the trilinear couplings  $A_{\l_i}$ to assume
small values. Such small values can be obtained through higher
dimensional operators involving SUSY and $U(1)$ symmetry breaking 
spurion fields \cite{Marshall:2009bk}. For instance, introducing a
superfield $\widehat{X}$ with charge $-1/2$ under $U(1)$ and with
$\langle X \rangle = \t^2 F + \sqrt{F}$  (see, e.g.,
\cite{ArkaniHamed:2000bq} for similar mechanisms), a trilinear  soft
term can originate from the operator $\frac{1}{M_{Pl}^2} \left|
\widehat{X}^2 \widehat{S} \widehat{H}_u^{\nu} \cdot  \widehat{H}_d
\right|_F \sim \frac{F^{3/2}}{M_{Pl}^2} S H_u^{\nu} \cdot H_d$.  If
$F=m_I^2$ with $m_I \simeq \sqrt{v M_{Pl}}$ an intermediate scale where
supersymmetry  is broken, then $\frac{F^{3/2}}{M_{Pl}^2} \sim 10^{-7}
\GeV$, while the corresponding  term in the superpotential is suppressed
by several orders of magnitude.

The resulting vevs for the $H_u^{\nu}$, $H_d^{\nu}$ fields have the form
\cite{Ma:2000cc}
\begin{equation} \label{eq:3}
 v_u^\nu \simeq \frac{A_{\l_2} s}{m^2_{H_u^\nu}}\, v_d \quad \textrm{
and } \quad 
 v_d^\nu \simeq \frac{A_{\l_1} s}{m^2_{H_d^\nu}}\, v_u,
\end{equation}
respectively. Taking $A_{\l_1} s \simeq A_{\l_2} s \sim 10^{-5} \GeV^2$
and assuming soft masses $m_{H_d^\nu} \sim m_{H_d^\nu} \sim
\mathcal{O}(1) \TeV$ (see below), then $v_u^\nu \simeq v_d^\nu \sim
\eV$. Hence, the first extra term in the superpotential  \eqref{eq:1}
will generate Dirac neutrino masses of the correct order for $y_\nu
\sim \mathcal{O}(1)$ \cite{Gabriel:2006ns,Davidson:2009ha}.

The mass squared matrix of the sneutrinos, neglecting flavor indices, reads
in the basis $\left( \wt{\nu}_L,\,\wt{\nu}_R \right)$
\begin{equation} \label{eq:4}
 \mathcal{M}_{\wt{\nu}}^2 = \left(
 \begin{array}{cc}
  y_\nu^2 {v_u^\nu}^2+\frac{1}{2}g^2(v_d^2-v_u^2)+m_{\nu_L}^2 & y_\nu
v_u^\nu (\l_\nu s +  A_\nu) \\
  & y_\nu^2 {v_u^\nu}^2 + m_{\nu_R}^2
 \end{array} \right).
\end{equation}
Taking into account the small value of $v_u^\nu$, this matrix can be
approximated by the  diagonal form
\begin{equation} \label{eq:5}
 \mathcal{M}_{\wt{\nu}}^2 \simeq \textrm{diag} \left[ \frac{1}{2}g^2 
(v_d^2-v_u^2)+m_{\nu_L}^2,\, m_{\nu_R}^2 \right].
\end{equation}
We note that the mixing between the different sneutrino flavors in the
right-handed  sector is small, since it is proportional to the vev of
the $\nu$-Higgs provided  that $m_{\nu_R^i} \gg v_u^\nu$ and flavour
diagonal.

The $\nu$-Higgses form two nearly degenerate SU(2) doublets. (Since
$U(1)$ is not spontaneously  broken, there are no Goldstone bosons.)
These additional fields mix very weakly  with the standard Higgs fields
due to their small vevs; in the following we will  consider the new
Higgs fields completely unmixed.

The mass matrices in the neutral sector are in the basis
$\left( H_u^\nu,\, H_d^\nu \right)$  \cite{Aranda:2000zf}
\begin{equation} \label{eq:6}
 \mathcal{M}_{H^\nu}^2 = \left(
 \begin{array}{cc}
   \l_\nu^2 s^2 - \frac{1}{2}g^2 (v_d^2-v_u^2) + m_{H_u^\nu}^2 & \pm \l_\nu (\l v_u v_d - 
\k s^2 + A_{\l_\nu} s) \\
   & \l_\nu^2 s^2 + \frac{1}{2}g^2 (v_d^2-v_u^2) + m_{H_d^\nu}^2
 \end{array} \right)
\end{equation}
with plus (minus) signs in the off-diagonal element for the scalar
(pseudoscalar), and
\begin{equation} \label{eq:7}
 \mathcal{M}_{{H^\nu}^+}^2 = \left(
 \begin{array}{cc}
   \l_\nu^2 s^2 + \frac{1}{2}g^2 (v_d^2-v_u^2)\cos{2\t_W} + m_{H_u^\nu}^2 & \l_\nu 
(\l v_u v_d - \k s^2 + A_{\l_\nu} s) \\
   & \l_\nu^2 s^2 - \frac{1}{2}g^2 (v_d^2-v_u^2)\cos{2\t_W} + m_{H_d^\nu}^2
 \end{array} \right)
\end{equation}
in the charged $\nu$-Higgs sector. The neutral and charged
$\nu$-higgsinos, forming Dirac fermions with masses 
$\mu^\prime = \l_\nu s$, are also practically unmixed with the  neutralinos and
the charginos of the NMSSM.

\subsection{Constraints from lepton flavour violation and BBN}
The charged Higgs ${H^\nu}^+$ mediates the decay of the muon at one
loop with a branching  ratio \cite{Fukuyama:2008sz}
\begin{equation} \label{eq:8}
 \textrm{BR}\left(\mu \rightarrow e\gamma \right) = 
 \frac{\a_{EM}}{24\pi}\left(\frac{v}{v_\nu m_{{H^\nu}^+}}\right)^4
 \left|\sum_j m_j^2 U_{ej}^* U_{\mu j} \right|^2,
\end{equation}
where $U$ is the Pontecorvo--Maki--Nakagawa--Sakata (PMNS) matrix
defined by  $ \left| \nu_l \right\rangle = \sum_{j=1}^3{U_{lj}^*} \left|
\nu_j \right\rangle $, with $l=e,\,\mu,\,\tau$ and $j=1,\,2,\,3$
corresponding to the three mass eigenstates. The unitarity of the PMNS
matrix allows to replace the sum in eq.\;\eqref{eq:8} by $\sum_j m_j^2
U_{ej}^* U_{\mu j}=-\Delta m_{21}^2 U_{e1}^* U_{\mu 1} + \Delta
m_{32}^2  U_{e3}^* U_{\mu 3}$, where the mass squared differences are
defined by $\Delta m_{ij}^2  \equiv m_i^2 - m_j^2$ and $m_i$ are the
neutrino mass eigenvalues. Using the upper $90\% \textrm{ C.L.}$ limit
\begin{equation} \label{eq:9}
 \textrm{BR}\left(\mu \rightarrow e\g \right) < 5.7 \cdot 10^{-13}
\end{equation}
from the MEG experiment \cite{Adam:2013mnn} gives a lower
bound on the charged $\nu$-Higgs mass,
\begin{equation} \label{eq:10}
 m_{{H^\nu}^+} \gtrsim \left( \frac{1\eV}{v_\nu} \right) 300\GeV,
\end{equation}
where we have used the standard values for  $\Delta m_{21}^2$, $\Delta
m_{32}^2$ given in \cite{Beringer:1900zz} and the elements  of the PMNS
matrix.

The additional degrees of freedom of the Dirac neutrinos contribute to
the energy density  and therefore to the expansion rate of the
universe during Big Bang Nucleo\-synthesis  (BBN). The abundance of
${^4}\textrm{He}$ emerging from BBN depends on the Hubble  expansion
rate when processes like $e^- + p \leftrightarrow n + \nu_e$ and $e^+ +
n  \leftrightarrow p + \bar{\nu}_e$ were in equilibrium, since
practically all the  remaining neutrons (after these processes went out
of equilibrium) were incorporated  in helium nuclei. The larger the
Hubble rate, the faster (at a higher temperature)  they are going out of
equilibrium, resulting in a larger abundance of neutrons\footnote{The
equilibrium density of neutrons falls as $n_n^{eq} \sim 
\exp(-\frac{\Delta m}{T}) n_p^{eq}$ with $\Delta m \equiv m_n - m_p$ the
mass  difference between neutron and proton.}.

In the epoch just before nucleo\-synthesis 
photons, electrons and left-handed neutrinos were in
equilibrium\footnote{The neutrinos decouple from the
thermal plasma at a temperature $T_d  \gtrsim 1\MeV$ when $H$ becomes
larger than the rate of the processes  $\nu+\bar{\nu}\rightarrow e^+ +
e^-$. Nevertheless, even after their decoupling but before the
decoupling of the electrons, their temperature is the same as
the one of photons since both are decreasing at the same
rate.} at a common temperature $T_{\g,n}$ (henceforth, the subscript
$n$ of temperatures  will denote  the epoch just before
nucleo\-synthesis).
Right-handed neutrinos remained in equilibrium as long as the processes
$\nu_R + \nu_R \leftrightarrow l+l$ (with $l$ a charged lepton),
mediated by the charged  $\nu$-Higgses, were fast enough. However, even
when the right-handed neutrinos go out of equilibrium at a
temperature $T_{R,d}$ (where the subscript $d$ stands for decoupling), 
they continue to contribute to the total energy density of  the universe
with their own temperature $T_R$ that is redshifting. 

Usually, the helium abundance is parametrized by the effective number of
degrees of  freedom $N_{eff}$ during BBN. Recently, Planck constrained
this quantity to  $N_{eff} = 3.30 \pm 0.27$ at $68\%$ C.L.
\cite{Ade:2013zuv}. In the following we determine the lowest temperature
$T_{R,d}$ at which the right-handed  neutrinos can decouple without
$N_{eff}$ exceeding the above limit, and subsequently we  derive the
necessary condition on the $\nu_R + \nu_R \leftrightarrow l+l$ rate for
this to occur.

Writing the energy density in the form
\begin{equation} \label{eq:11}
 \rho_n = \frac{\pi^2}{30}\left[ g_\g + \frac{7}{8}\left(g_e+N_{eff}\,g_\nu\right)\right] 
T_{\g,n}^4,
\end{equation}
$N_{eff}$ is defined as $N_{eff} \equiv n_L+n_R
\left(\frac{T_{R,n}}{T_{\g,n}}\right)^4$ with $n_L$ ($n_R$)  the number
of left- (right-) handed neutrino generations. Taking $n_L=n_R=3$ we
write $N_{eff}=3+\Delta N_\nu$ with
\begin{equation}\label{eq:12}
\Delta N_\nu \equiv 3 \left(\frac{T_{R,n}}{T_{\g,n}}\right)^4.
\end{equation}

Applying entropy conservation 
separately for the decoupled species and the thermal bath
\cite{Steigman:1979xp} one finds $\frac{T_{R,n}}{T_{\g,n}} = \left(
\frac{43}{4 g(T_{R,d})} \right)^{1/3}$, where $g(T_{R,d})$ is the number
of degrees of freedom when the right-handed neutrinos  decouple.
Substituting the last relation into \eqref{eq:12}, we obtain for the
relation between the maximally allowed value of $\Delta N_\nu^{max}$ and
the temperature  at which  the right-handed neutrinos went out of
equilibrium
\begin{equation} \label{eq:13}
 g(T_{d,R}) \geq \frac{43}{4} \left( \frac{3}{\Delta N_\nu^{max}}
\right)^{3/4}.
\end{equation}
For $\Delta N_\nu^{max} \lsim 0.57$ at $1\s$, $g(T_{R,d}) \geq 37.35$.
This means that decoupling should have occurred before the
quark-hadron phase transition when  $g = 51.25$ (just after the
transition the number of degrees of freedom was $g=17.25$). Assuming
that the QCD confinement temperature is  roughly $T_c \simeq 200 \MeV$
\cite{Karsch:2000ps} leads to the inequality  $T_{R,d} \gsim 200 \MeV$.

Taking into account the approximate decoupling condition $n(T_d) \langle
\s v  \rangle(T_d) = H(T_d)$, one finds that the ratio of the decoupling
temperatures of  right- and left-handed neutrinos is
\begin{equation} \label{eq:14}
 \left( \frac{T_{R,d}}{T_{L,d}} \right)^3 =
\sqrt{\frac{g(T_{L,d})}{g(T_{R,d})}}  \frac{\s_L}{\s_R},
\end{equation}
with $\s_L$ and $\s_R$ the cross sections of the processes that were
keeping left- and  right-handed neutrinos, respectively, in equilibrium.
Using $\frac{\s_L}{\s_R} = \left(
\frac{2\sqrt{2}m_{{H^\nu}^+}}{y_\nu^{li} v_u | U_{li} |} \right)^4$
\cite{Davidson:2009ha}, where $U_{li}$ are again the elements of the
PMNS mixing matrix, leads to the following bound on the charged Higgs
mass and the  couplings $y_\nu^{li}$
\begin{equation} \label{eq:15}
 \frac{m_{{H^\nu}^+}}{y_\nu^{li}} \gtrsim 3\TeV.
\end{equation}
As we will explain later, the couplings $y_\nu^{li}$ cannot be very
small, and as a  consequence $m_{{H^\nu}^+}$ has to be relatively large.

\section{Right-handed sneutrinos as ADM} \label{DM}

In the following we study more closely the r\^ole of right-handed
neutrinos as ADM. We  will use the notation $N \equiv
{{}\wt{\nu}_R^c}_1$ with the index $1$ denoting the  lightest among the
three right-handed neutrinos. Its mass $m_N$ is essentially its soft
Susy breaking mass, and we  safely assume that it is a pure state since its
left-handed component is negligibly small.

\subsection{Asymmetry from sphaleron processes and the ADM mass}\label{LeptonNumber}

Since the sneutrinos carry a conserved charge (lepton number), it is
possible that their relic density is not determined by the thermal
mechanism but by their asymmetry. The asymmetry was related to the
baryon asymmetry through equilibrium processes in the early universe. These
allow to estimate the relation
between the two asymmetries and, ultimately, to determine the mass range
of the (right-handed sneutrino) DM that will provide the correct abundance.

If the $N,\,N^*$ annihilation is strong enough such that the less
frequent species has been completely eliminated, the remaining abundance
is the product of the charge density $\eta_N \equiv | n_N -
n_{N^*}|$ times its mass $m_N$ ($n$ denotes the number density). The
relation between the DM relic density $\Omega_N$ and the baryonic relic
density $\Omega_b$ is
\begin{equation} \label{eq:16}
 \Omega_N = \frac{\eta_N}{B} \frac{m_N}{m_p}\, \Omega_b,
\end{equation}
where $m_p$ is the proton mass, which gives the desired result if
$\eta_N$ is of same order of magnitude as the baryon charge density $B$.

The charge density of a particle $X$ in kinetic equilibrium as a
function of the temperature can be written as
\begin{equation} \label{eq:17}
 \eta_X(T) = \frac{T^3}{6} g_X k(x) \frac{\mu_X}{T},
\end{equation}
where we have assumed that $\mu_X / T \ll 1$, and $\mu_X$ is the
chemical potential of the species X. $g_X$ is the number of internal
degrees of freedom of the particle $X$, we defined $x\equiv
\frac{m_X}{T}$, and
\begin{equation} \label{eq:18}
 k(x) = \frac{6}{\pi^2} \int_x^\infty{\di y \,y\sqrt{y^2-x^2}
\frac{\e^y}{(\e^y\pm 1)^2}}.
\end{equation}
In the above integral, the {plus} ({minus}) sign holds for
fermions (bosons). In the ultra-relativistic limit $x \ll 1$, $k$ takes
the values $1$ for fermions and $2$ for bosons, while in the opposite
limit $x \gg 1$ it vanishes in both cases. 

A sneutrino asymmetry can originate from primordial asymmetries in the
baryonic or leptonic sectors.
Although we will be agnostic about the
exact mechanism that created these primordial asymmetries, the fact that
certain processes were in equilibrium in the early universe can be used
to relate $\eta_N$ to the baryon asymmetry $B$. We note
that common mechanisms for thermal leptogenesis would not work in the
present framework since the violation of lepton number is far too small
(see the next section). 
However, other known
mechanisms are possible, such as the Affleck-Dine mechanism \cite{Affleck:1984fy}.

In the absence of lepton number violating processes other than
electroweak spha-lerons, $\sum_{i=1}^3{\left(B/3-L_i\right)}$ is
conserved. The relatively large Yukawa coupling constants of the
neutrinos assure not only the equilibrium of the right-handed neutrinos
with the thermal bath in the early universe, but also rapid flavor
changing processes. As a result, lepton flavor equilibrium had been
established and $B-L=\sum_{i=1}^3{\left(B/3-L_i\right)}$ is conserved.
However, the sphaleron processes were still violating $B+L$. We are
going to consider two cases \cite{Harvey:1990qw}. In the first case we
will assume that sphaleron processes were rapid only above the
electroweak phase transition (EWPT), e.g. if the EWPT was strongly first
order. In the second case, we will allow the sphaleron processes to
violate $B+L$ also below the transition, until they went out of
equilibrium because of the expansion of the universe.

\begin{figure}[tb]
 \begin{center}
  \includegraphics[scale=0.7,angle=0]{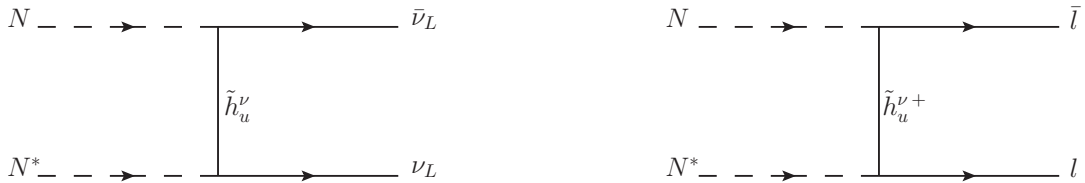}
 \end{center}
 \caption{Annihilation diagrams of right-handed sneutrinos $N$, $N^*$
into neutrinos and charged leptons.}
 \label{fig:a}
\end{figure}

In the first case we proceed along the lines of \cite{Chung:2008gv},
where one can find a complete list of the equilibrium reactions in the
MSSM and the relations between the chemical potentials. The reactions
specific to the present model lead to the following equilibrium
relations which have to be added to this list:
\begin{eqnarray} \label{eq:19}
 \mu_{L^i}+\mu_H \= \mu_{\nu^i},  \quad \mu_H =
\mu_{H_u^\nu}=\mu_{H_d^\nu}, \nn \\
 \mu_{\widetilde{L}^i}+\mu_H \= \mu_{\widetilde{\nu}_i}, \quad
\mu_{\widetilde{L}^i}+\mu_{\widetilde{H}}=\mu_{\nu_i}, \quad
  \mu_{L^i}+\mu_{\widetilde{H}}=\mu_{\widetilde{\nu}_i},
\end{eqnarray}
where we have used the notation of \cite{Chung:2008gv}, i.e. $\mu_{L^i}$
is the chemical potential of the left-handed leptons, $i$ is the flavor
index, $\mu_\nu$ is the chemical potential of the right-handed neutrinos
and \emph{tilde} stands for the supersymmetric particles. The sneutrinos
share the chemical potential with the neutrinos through the
equilibrium of processes such as those of \figref{fig:a}.

Eliminating the chemical potentials using the sphaleron equilibrium
relation and the fact that the total hypercharge of the universe
vanishes, we can calculate the baryon charge $B$ and the DM leptonic
charge $\eta_N$ as functions of the conserved difference $B-L$. We
assume all the supersymmetric particles except the sleptons much heavier
than the EWPT critical temperature $T_c$, and take massless SM
particles. First, assuming light right-handed sneutrinos and masses $\sim
2T_c$ for the other sleptons, $B$ and $\eta_N$ are related by
\begin{subequations} \label{eq:20}
\begin{equation} \label{eq:20a}
 B \simeq 0.14\, (B-L) \quad \textrm{ and } \quad \eta_N \simeq 0.10\,
(B-L), \quad \textrm{ (light sleptons)}
\end{equation}
while for large slepton masses
\begin{equation} \label{eq:20b}
 B \simeq 0.18\, (B-L) \quad \textrm{ and } \quad \eta_N \simeq 0.12\,
(B-L). \quad \textrm{ (heavy sleptons)}
\end{equation}
\end{subequations}
The DM mass, using \eqref{eq:16} and $\Omega_N / \Omega_b \simeq 5.44$
\cite{Ade:2013zuv}, has to be $m_N \sim 7.1 \,-\, 7.6\GeV$ (the smaller
value corresponding to light sleptons).

In case the process induced by electroweak sphalerons were rapid also
below the EWPT, the relations between the chemical potential are
altered. First, due to the vacuum condensate of the neutral Higgs
bosons, their chemical potentials have to vanish. However, the total
hypercharge has no longer to be zero since $SU(2)_L$ has been broken,
resulting in a non-zero chemical potential for the $W$ bosons. Generalizing the
SM equilibrium processes of \cite{Harvey:1990qw} we find, considering
all the supersymmetric particles (except for the right-handed
sneutrinos) as heavy,
\begin{subequations} \label{eq:21}
\begin{equation} \label{eq:21a}
 B \simeq 0.18\, (B-L) \quad \textrm{ and } \quad \eta_N \simeq 0.10\,
(B-L). \quad \textrm{ (heavy SUSY paricles)}
\end{equation}
The resulting DM mass in this case has to be $m_N \simeq 9.2 \GeV$.
However, allowing the left-handed sneutrinos to be light (with mass
around the temperature at which the sphaleron processes went out of
equilibrium), the value for the ratio $B/\eta_N$ becomes maximal:
\begin{equation} \label{eq:21b}
 B \simeq 0.31\, (B-L) \quad \textrm{ and } \quad \eta_N \simeq 0.07\,
(B-L). \quad \textrm{ (light LH sneutrinos)}
\end{equation}
\end{subequations}
In this case the DM mass has to be larger, $m_N \simeq 23 \GeV$.

Summarizing, depending on the sparticle spectrum and the nature of the
EWPT, the DM mass can roughly be in the range
\begin{equation} \label{eq:22}
 m_N \sim 7\GeV \,-\, 23 \GeV.
\end{equation}
The lowest value corresponds to light sleptons and a first order EWPT
that terminated the sphaleron processes, while for the highest value
the sphaleron processes have to continue to be in equilibrium for a
short time after the EWPT and the left-handed sneutrinos have to be
relatively light.

\subsection{Constraints from oscillations, self and pair annihilation}
\label{SneutrinoADM}

In order that the current DM density to be determined by its
asymmetry, a number of conditions have to be fulfilled. First, the
annihilation of DM particles with antiparticles has to be strong enough
so that one of them is completely depleted. However, in many cases, it
is possible for a particle to oscillate into its antiparticle and vice
versa. These oscillations, if rapid enough, might lead to a continuous
repopulation of the depleted particles. As a result, however strong the
pair annihilation cross section is, the antiparticles (or the particles)
are never exhausted and, finally, the thermal mechanism is responsible
for the relic density. Furthermore, if self-annihilation\footnote{the
particle--particle or the antiparticle--antiparticle annihilation.} of
DM particles occurs before the DM particles become non-relativistic,
their asymmetry decreases rapidly due to this annihilation. If the
self-annihilation does not freeze-out sufficiently fast, the thermal
mechanism takes over again since there is no asymmetry left after the
particle--antiparticle annihilation freeze-out. We will show that the
sneutrino DM considered here can fulfill all these criteria for
a successful asymmetric DM candidate.

Quantum mechanical oscillations occur between $N$ and $N^*$ if they are
not the mass eigenstates. Then the rate of $N$ -- $N^*$ conversion
is approximately given by the mass difference $\d m$ of the two
eigenstates. The conversion starts to be significant only at times
larger than $\d m^{-1}$ or, expressed in terms of the temperature $T$ of
the universe, for $T \lsim T(\d m)$ given by \cite{Buckley:2011ye}
\begin{equation} \label{eq:23}
 T(\d m) \sim \left( \frac{g_*^{1/2}}{h_{eff}} \sqrt{\frac{45}{4\pi^3}}
M_{Pl}\, \d m \right)^{1/2},
\end{equation}
where $M_{Pl}$ is the Planck mass and $g_*$ and $h_{eff}$ are effective
degrees of freedom (see, e.g., \cite{Ellwanger:2012yg} for
exact definitions).

A mass split appears if there exists a lepton number violating Majorana
mass term $m_M$; if $m_M \ll m_D$ ($m_D$ is the Dirac mass), the mass
split can be written as $\d m \simeq \frac{m_M^2}{m_D}$. 
The operator $\frac{1}{M_{Pl}^4}
\left| X^4 S N^2 \right|_F \nonumber$, with $X$ the superfield spurion
whose vev brakes the $U(1)$ symmetry, induces a tiny Majorana mass squared of the
order $m_M^2 \simeq 10^{-32} \GeV^2$, which corresponds to a mass
difference $\d m \sim 10^{-33} \GeV$ for $m_D \sim 10\GeV$. With such a
small value for the Majorana mass, the oscillations start very late in
the history of the universe (see \eqref{eq:23}), much later than the DM
freeze-out (at $T\sim 1\GeV$), and do not affect the final DM density.

Upper bounds on the self-annihilation cross section have been derived in
\cite{Ellwanger:2012yg}. If the cross section does not obey these
bounds, the asymmetry falls rapidly. In order that at least $90\%$ of
the asymmetry survives, the decoupling of self-annihilation should
happen before $x \equiv m_N/T\sim 5$ (we recall that the decoupling for
WIMPs occurs at $x \sim 20$ -- $30$). However, in our case the possible
annihilation of right-handed sneutrinos into two neutrinos through t- or
u-channel exchange of neutral $\nu$-higgsinos is impossible due to the
Dirac nature of the $\nu$-higgsinos. Furthermore, the left-handed
components in $N$ and $N^*$ are sufficiently small, since they are
induced only by the off-diagonal element of the mass matrix \eqref{eq:4}
and hence many orders of magnitude below the bound of
\cite{Ellwanger:2012yg}. Consequently, the self-annihilation cross
sections of $N$ or $N^*$ are sufficiently small.

Now that we have shown that the asymmetry does not get destroyed by
oscillations or self annihilations, the condition that remains to be
satisfied is a sufficiently strong $N,\,N^*$ pair annihilation so that
only the asymmetry survives as relic density. The dominant annihilation
channels of right-handed sneutrinos are the annihilation into neutrinos
and charged leptons (\figref{fig:a}). The former proceeds through a
t-channel neutral $\nu$-higgsino exchange, the latter by charged
$\nu$-higgsino exchange. The thermal average of the cross section of
these processes times velocity can be written as (to leading order in
$x^{-1}$)
\begin{equation} \label{eq:24}
 \langle \s v \rangle \simeq f \, \frac{y_\nu^4}{8\pi}
\frac{m_N^2}{(m_N^2+{\mu^\prime}^2)^2} x^{-1},
\end{equation}
where the factor $f=18$ counts the number of final states (9 
neutrinos and 9 charged leptons) and we have assumed a common value
$y_\nu$ for the coupling constants $y_\nu^{li}$. The s-wave contribution
is helicity suppressed and can be neglected (see also
\cite{Lindner:2010rr}).

In the usual symmetric DM case, the thermally averaged cross section
during freeze out has to be of the order of the so-called thermal cross
section, roughly given by
\cite{Jungman:1995df}
\begin{equation} \label{eq:25}
 \langle \s v \rangle_{th} \simeq \frac{3\cdot 10^{-27}}{\Omega_{DM}h^2}
			   \;\textrm{cm}^3\textrm{s}^{-1} \simeq 3\cdot
			   10^{-26} \;\textrm{cm}^3\,\textrm{s}^{-1}
\end{equation}
from considerations using entropy conservation. In the asymmetric DM
scenarios, the pair annihilation cross section must be equal to or
larger than the thermal cross section; even if the cross section is
much larger than its thermal value, the final density remains constant
since annihilations become impossible due to the lack of $N$ or
$N^*$. Examining eq.\,\eqref{eq:24} at fixed $m_N$, the cross section
decreases with increasing mass $\mu^\prime$ of the $\nu$-higgsino.

\begin{figure}[tb]
 \begin{center}
  \includegraphics[scale=1.1,angle=0]{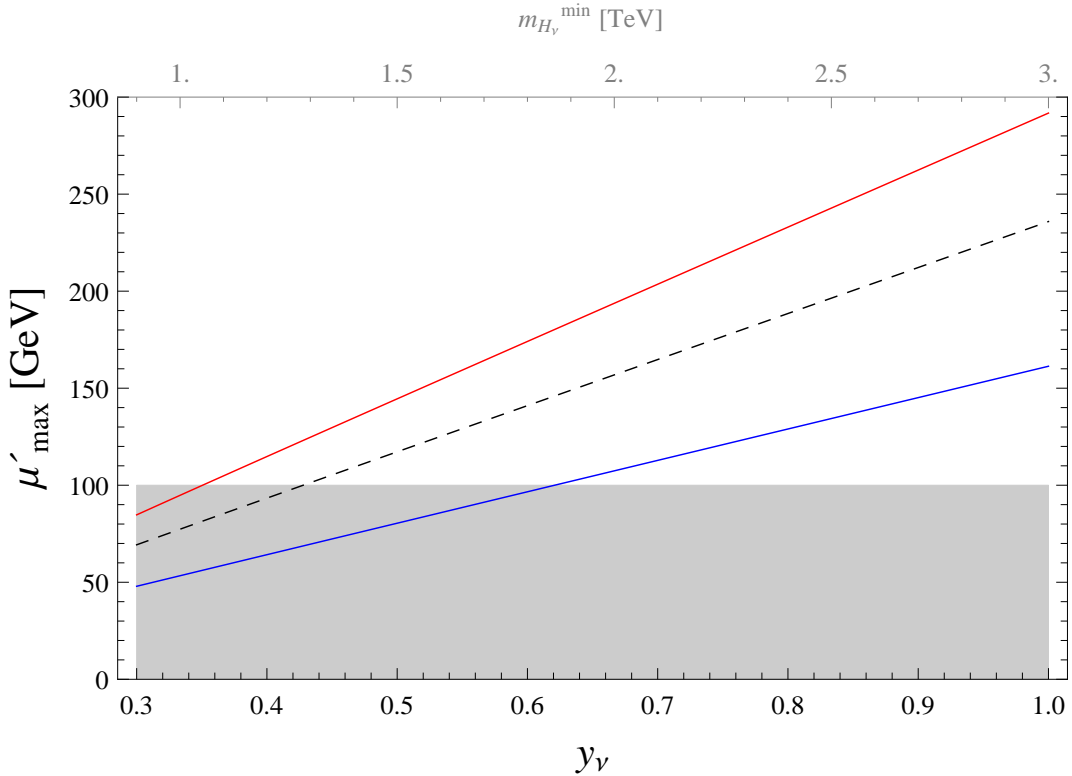}
 \end{center}
 \caption{The largest allowed $\nu$-higgsino mass as function of the
coupling $y_\nu$ such that right-handed sneutrinos with mass $7\GeV$
(red --upper-- line), $15\GeV$ (dashed line) or $23\GeV$ (blue --lower--
line) have a large enough pair annihilation cross section such that their
relic density is determined by their asymmetry.  The corresponding lower
limit on the lightest $\nu$-Higgs mass, derived from eq.
\!\eqref{eq:15}, is indicated along the upper axis. The shaded area is
excluded by chargino searches at LEP (e.g. \cite{Abbiendi:2003ji}).} \label{fig:c}
\end{figure}

\figref{fig:c} shows the maximal value of the $\nu$-higgsino mass as
function of the coupling constant $y_\nu$ for a sufficiently large
annihilation cross section. The corresponding minimal allowed mass of
the lightest scalar charged $\nu$-Higgs (from the condition that
$N_{eff}$ is within the $1\sigma$ region determined by Planck) is shown
along the upper axis. For a light right-handed sneutrino mass ($7\GeV$)
the coupling constant has to be relatively large, $y_\nu \gsim 0.6$,
which requires $m_{H_1^\nu} \gsim 1.8\TeV$. Smaller values of $y_\nu$
(and lower bounds on the scalar $\nu$-Higgs mass) are allowed for
heavier DM. For $m_N = 23 \GeV$, the smallest allowed value for the
$\nu$-Higgs mass is $\sim 1 \TeV$.

We see that the right-handed sneutrino can have a relic density
determined by its asymmetry if the $\nu$-higgsino is relatively light,
while the scalar $\nu$-Higgs should be heavy. In terms of the coupling 
$\l_\nu$ in the NMSSM Lagrangian and the singlet vev $s$,
 the $\nu$-higgsino mass is given by  $\mu^\prime= \l_\nu\, s$ and hence
small for small $\l_\nu$,
whereas a heavy charged $\nu$-Higgs requires a large soft Susy breaking
mass.

\subsection{ADM Detection: prospects and constraints} 
\label{detection}

Upper bounds on the ADM-nucleon scattering cross section originate from
both the direct detection and observations of old neutron stars.
Concerning the latter, if the cross section is too large, the
accumulation of asymmetric DM inside the neutron stars can form a black
hole which would potentially destroy the star. This is a specific
feature of asymmetric DM, since in the common symmetric case the
annihilation of DM with anti-DM prevents the accumulation.
For bosonic
asymmetric DM in the mass range $5\GeV\lsim m_{DM} \lsim 16 \GeV$,
nucleon--DM cross sections $\s \gsim 10^{-50}$ are excluded
\cite{Kouvaris:2011fi,McDermott:2011jp}. However, the value of the cross
section depends on the parameter space, particularly on the value of
$A_\nu$ (see \cite{Choi:2012ap}), while for DM heavier than $\sim 16
\GeV$ there is no limit due to Hawking evaporation
\cite{Kouvaris:2011fi}, letting a completely unconstrained mass
range ($16\GeV \lsim m_{DM} \lsim 23\GeV$) for the ADM of this scenario.

Concerning direct detection,
since right-handed sneutrinos couple only to neutrinophilic Higgses,
there are no tree-level contributions to the scattering cross section of
$N$ off nuclei. However, as it was pointed out in
\cite{Choi:2012ap}, the contribution of Z exchange induced at one
loop (with left-handed sneutrinos and $\nu$-Higgses running on the loop)
may be significant. The value obtained in \cite{Choi:2012ap} 
($\sim 10^{-45}\,\textrm{cm}^3
\textrm{s}^{-1}$), assuming a relatively large value
for the trilinear soft coupling $A_\nu$, is at the lower bound of the
current experimental direct detection sensitivity for a DM mass around
$100\GeV$. However, for the mass range considered here
($\mathcal{O}(10)\GeV$), the upper limits on the scalar scattering cross
section are much higher. 

Concerning indirect detection, pure asymmetric DM does not
give rise to detectable signals due to the absence of either DM
particles or antiparticles. (The self-annihilation cross section
is required to be too small to generate measurable signals.)
However, the operators which break the $U(1)$ symmetry might also induce a
very small mass difference $\d m$ among the sneutrino and
anti-sneutrino eigenstates. Even though the induced sneutrino -- 
anti-sneutrino oscillations are slow
enough in order not to affect the relic density, it may have led to the
repopulation of the exhausted species if $\d m^{-1}$ is smaller than the
current age of the universe. This would be the case, e.g., for the value
$\d m \sim 10^{-33} \GeV$ obtained in the scenario sketched below
\eqref{eq:23}.

In case the exhausted species has been regenerated, the same
$N,\,N^*$ annihilation processes (\figref{fig:a}) that occurred in the
early universe may happen today in galactic regions of large DM density,
giving rise to leptonic charged cosmic rays and $\g$-rays. However, 
assuming that the excess of positrons observed amongst others by AMS-02
\cite{Aguilar:2013qda} originates from
astrophysical sources, it constitutes an insurmountable background making
the distinction of a potential DM signal from charged leptonic
rays difficult. Concerning the diffuse photon radiation, we recall that
the s-wave annihilation of $N,\,N^*$ is helicity suppressed. The low
present-day velocity of DM particles leads to a low $\s v$, evading the
bounds set by the Fermi collaboration \cite{Mazziotta:2012ux}. Finally,
as pointed out in \cite{Choi:2013fva}, $N,\,N^*$ annihilation through a
box loop with sleptons and charged $\nu$-Higgs can give rise to a
monochromatic photon line with a large cross section proportional to
$\left( \frac{y_\nu A_\nu}{M_{H_\nu^+}} \right)^4$. The Fermi
bound for a DM mass of $\sim 10 \GeV$ is quite severe, $\langle \s v
\rangle_{\g\g} \lsim 5 \cdot 10^{-29} \,\textrm{cm}^3 \textrm{s}^{-1}$
\cite{Ackermann:2012qk}. However, taking $A_\nu$ of the order of the 
EW scale ($\sim
100 \GeV$), this bound is satisfied since $\langle \s v \rangle_{\g\g}
\lsim 10^{-29} \,\textrm{cm}^3 \textrm{s}^{-1}$.

\section{Summary and outlook} \label{conclusions}
In this paper we have presented an extension of the NMSSM introducing an
additional pair of Higgs doublets with small vevs, explaining the smallness
of neutrino masses
and, at the same time, the present day coincidence of DM and baryon
densities. The additional Higgses and the right-handed neutrinos are
charged under a new $U(1)$ symmetry which is explicitely broken by soft
SUSY breaking
terms. This symmetry forces the new Higgses to couple in the
superpotential only to right-handed neutrinos 
(so-called neutrinophilic Higgses). The neutrinos have Dirac masses 
which are
generated dynamically by the neutrinophilic Higgs vev and hence
naturally small.

We have shown that the right-handed sneutrinos can carry an asymmetry related to
the baryon asymmetry due to their conserved lepton number and
equilibrium processes in the early universe. They can maintain their
asymmetry at least until the  freeze-out of
sneutrino--anti-sneutrino annihilations. Therefore their
relic density is determined by their asymmetry and of the correct value
if their mass is $\mathcal{O}(10) \GeV$, provided that the coupling
constant $\l^\prime$ is small compared to $\l$. However, the bound on
the relativistic degrees of freedom during BBN set by the Planck
collaboration requires large soft breaking mass for the neutrinophilic
Higgs. At present this scenario satisfies constraints from
DM detection experiments.
Actually, the scattering cross section is too small in order to explain
possible excesses observed in the CDMS, DAMA, CoGENT and CRESST-II
experiments
\cite{Agnese:2013cvt,Bernabei:2008yi,Aalseth:2010vx,Angloher:2011uu}
in this mass range, which have been interpreted as
possible evidence of DM.
Still, direct detection is possible in the future once the
sensitivity in the lower mass range is improved.
On the other hand, neutrinoless double beta decay is impossible in this model and
a future observation of this process would rule out the current scenario.

\section*{Acknowledgments}
First and foremost, the author is grateful to U. Ellwanger for useful
discussions and valuable comments throughout this work.
He would also like to thank A. Abada and A. Vicente for discussions.
P.~M. acknowledges support from the Greek State Scholarship Foundation.

\bibliography{bib}{}

\end{document}